\documentclass{article}

\usepackage{arxiv}

\usepackage[utf8]{inputenc} 
\usepackage[T1]{fontenc}    
\usepackage{hyperref}       
\usepackage{url}            
\usepackage{booktabs}       
\usepackage{amsfonts}       
\usepackage{nicefrac}       
\usepackage{microtype}      
\usepackage{lipsum}		
\usepackage{graphicx}
\usepackage[square,sort,comma,numbers]{natbib}
\usepackage{doi}
\usepackage{float}
\usepackage{pdfpages}

\title{Astrotourism for Development: An Overview of Resources from the IAU Office of Astronomy for Development}

\author{ \href{https://orcid.org/0000-0002-9745-0504}{\includegraphics[scale=0.06]{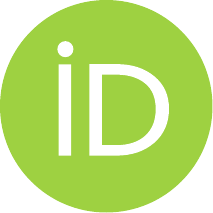}\hspace{1mm}Joyful E. Mdhluli}{ on behalf of the IAU Office of Astronomy for Development} \thanks{Visit our website, www.astrotourism.astro4dev.org} \\
	International Astronomical Union's Office of Astronomy for Development\\
	South African Astronomical Observatory\\
        Cape Town, South Africa\\
	\texttt{joy@astro4dev.org} \\
	\And
}

\date{}

\renewcommand{\shorttitle}{\it{Astrotourism Resources}}

\hypersetup{
pdftitle={Astrotourism for Development: An Overview of Resources from the IAU Office of Astronomy for Development},
pdfsubject={astro-ph.IM},
pdfauthor={Joyful E.~Mdhluli},
pdfkeywords={Astronomy, Sustainable development, Tourism, Astrotoursim},
}

\begin{document}
\maketitle

\begin{abstract}
Astrotourism has emerged as a powerful cross-sectoral tool to promote science education, sustainable economic development, and cultural exchange. Recognising its potential, the International Astronomical Union’s Office of Astronomy for Development (IAU OAD) has developed a suite of openly accessible resources to support individuals and institutions interested in implementing astrotourism initiatives globally. These resources also encourage individuals and existing businesses to broaden their offerings to include activities that use the night sky as a backdrop, such as food experiences, wellness practices, and cultural exploration. This paper offers a comprehensive summary of these resources, available on the OAD’s Astrotourism Portal, and situates them within the broader context of astronomy-for-development work. The paper is targeted at educators, policymakers, tourism operators, grassroots organisers, and entrepreneurs, providing guidance on how they can foster inclusive, locally grounded, and sustainable astrotourism efforts, particularly in under-resourced or emerging contexts.
\end{abstract}

\section{Introduction}

Astrotourism, broadly defined as travel and tourism experiences centred on astronomy - such as stargazing and visits to sites of astronomical heritage - but also includes activities where astronomy provides a unique backdrop, such as dining under the stars, has gained global traction as a form of nature based tourism. Combining scientific learning with natural and cultural appreciation, astrotourism offers unique opportunities for public participation, rural development, and environmental conservation.

The International Astronomical Union’s (IAU) Office of Astronomy for Development (OAD) works to promote the use of astronomy in addressing global issues encapsulated by the \href{https://sdgs.un.org/goals}{Sustainable Development Goals} (SDGs) \cite{sdgs} of the United Nations \cite{comment1}\cite{comment2}. Astrotourism is recognised by the OAD as one of its focus areas due to its ability to intersect multiple development goals while increasing science capital and promoting socioeconomic development in communities worldwide.

This paper presents an overview of key resources created by the OAD to support the global growth of astrotourism. These materials, hosted on the \href{https://astrotourism.astro4dev.org}{OAD Astrotourism Portal} \cite{astrotourism}, provide practical guidance, toolkits, and learning content aimed to enable anyone, from community leaders to tourism professionals, to initiate or improve astrotourism projects.

\section{Context and Motivation}
Traditionally, astrotourism has been a niche pursuit, centred on scientific exploration and specialised equipment, and targeted at a specific audience of science enthusiasts. However, this narrow focus has limited its potential to drive socioeconomic development in rural and/or remote communities.

The OAD's approach is a radical departure from the norm, re-imagining astrotourism as an extension of ecotourism and cultural tourism, unlocking the full potential of the night sky. Astrotourism, as a form of informal science outreach, combines accessible astronomy education with local cultural knowledge, enabling communities to engage visitors meaningfully bringing together their traditions, stories, and understanding of the stars, creating an enriching experience that connects science with culture and place.

Astrotourism offers a rare convergence of SDG goals: fostering interest in science, supporting environmental conservation (through dark sky preservation), stimulating local economies, and celebrating cultural and indigenous knowledge systems. This aligns directly with several SDGs, including:
\begin{itemize}
  \item \textbf{SDG 4}: Quality Education
  \item \textbf{SDG 8}: Decent Work and Economic Growth
  \item \textbf{SDG 11}: Sustainable Cities and Communities
  \item \textbf{SDG 17}: Partnerships for the Goals
\end{itemize}

\begin{figure}[H]
    \centering
    \includegraphics[width=0.2\textwidth]{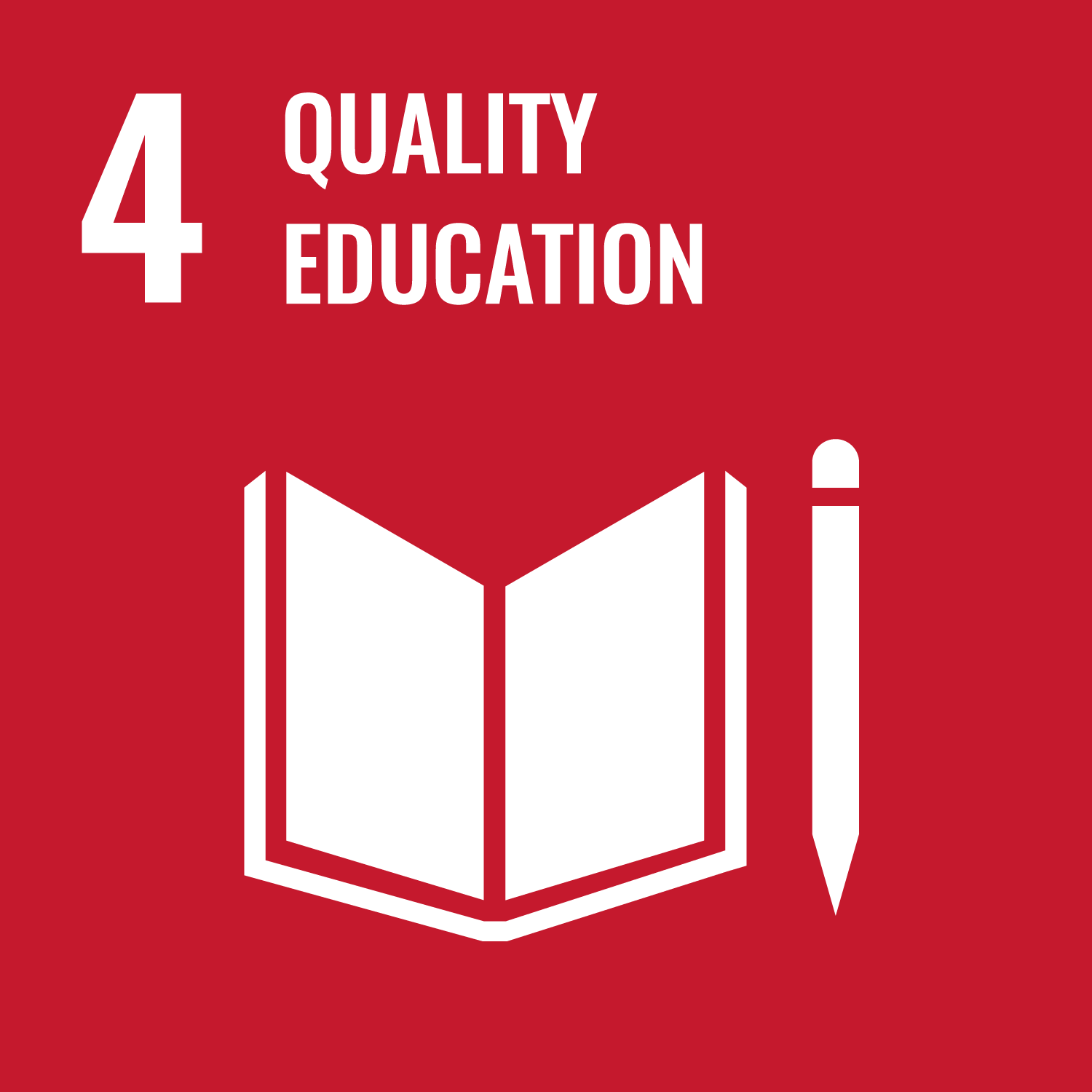}
    \includegraphics[width=0.2\textwidth]{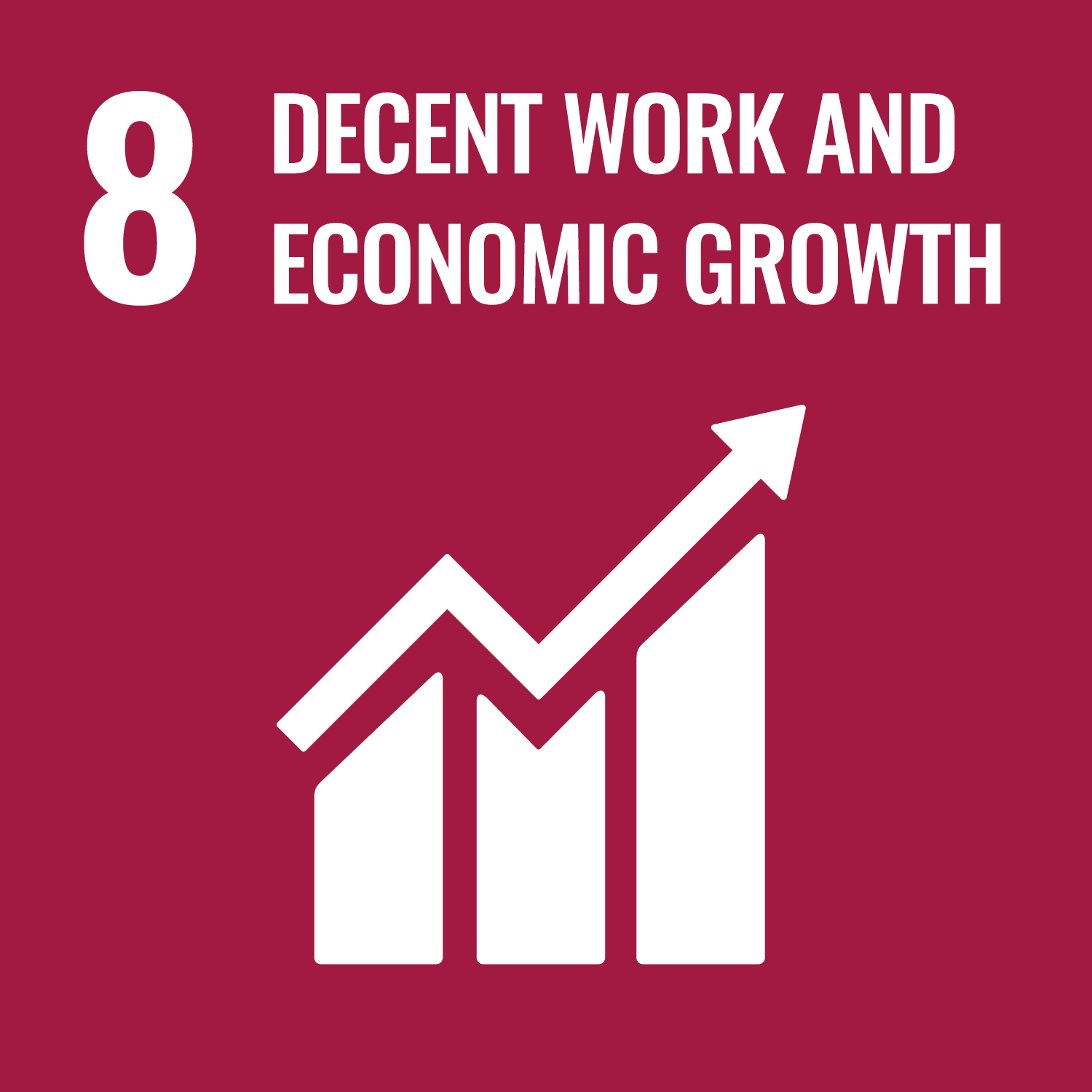}
    \includegraphics[width=0.2\textwidth]{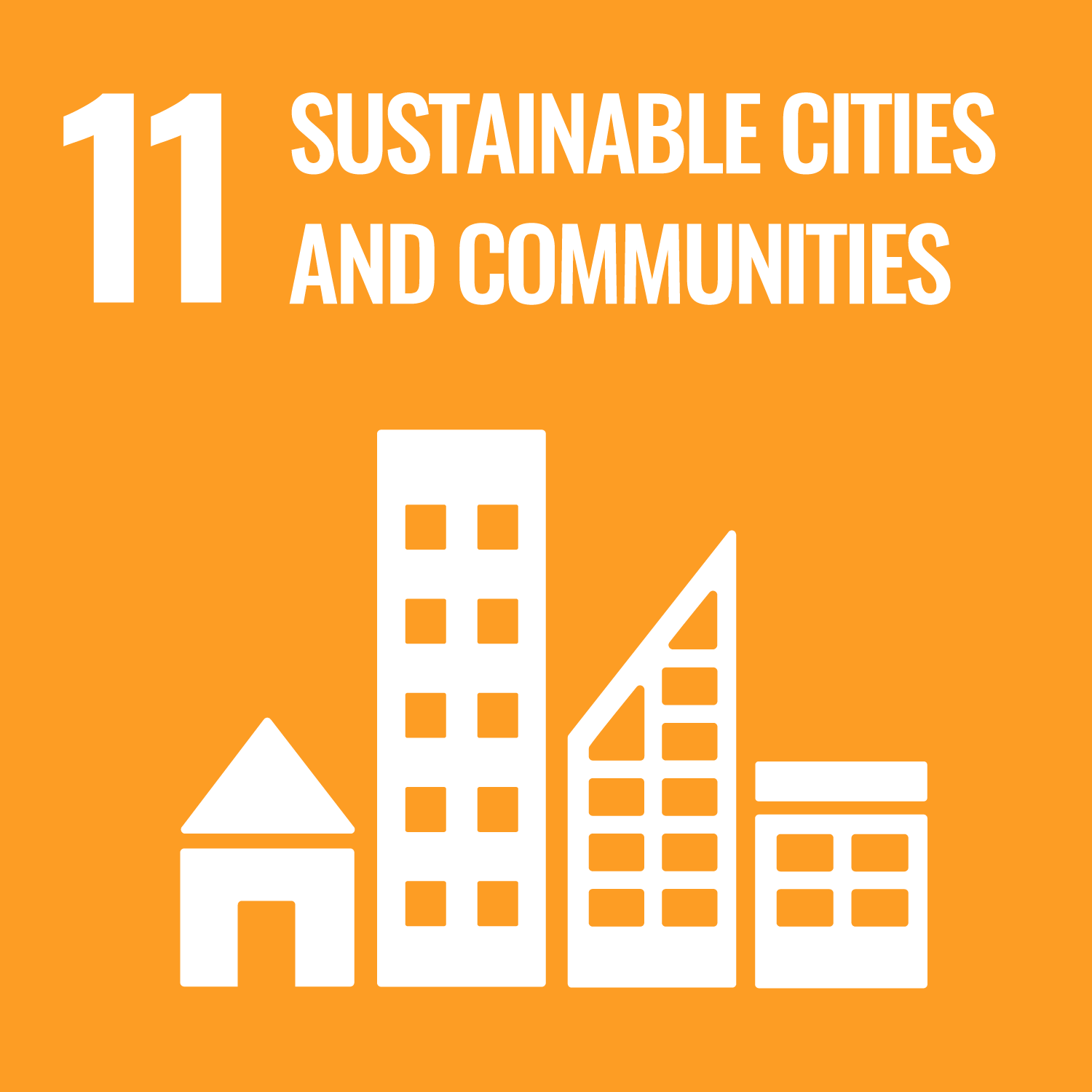}
    \includegraphics[width=0.2\textwidth]{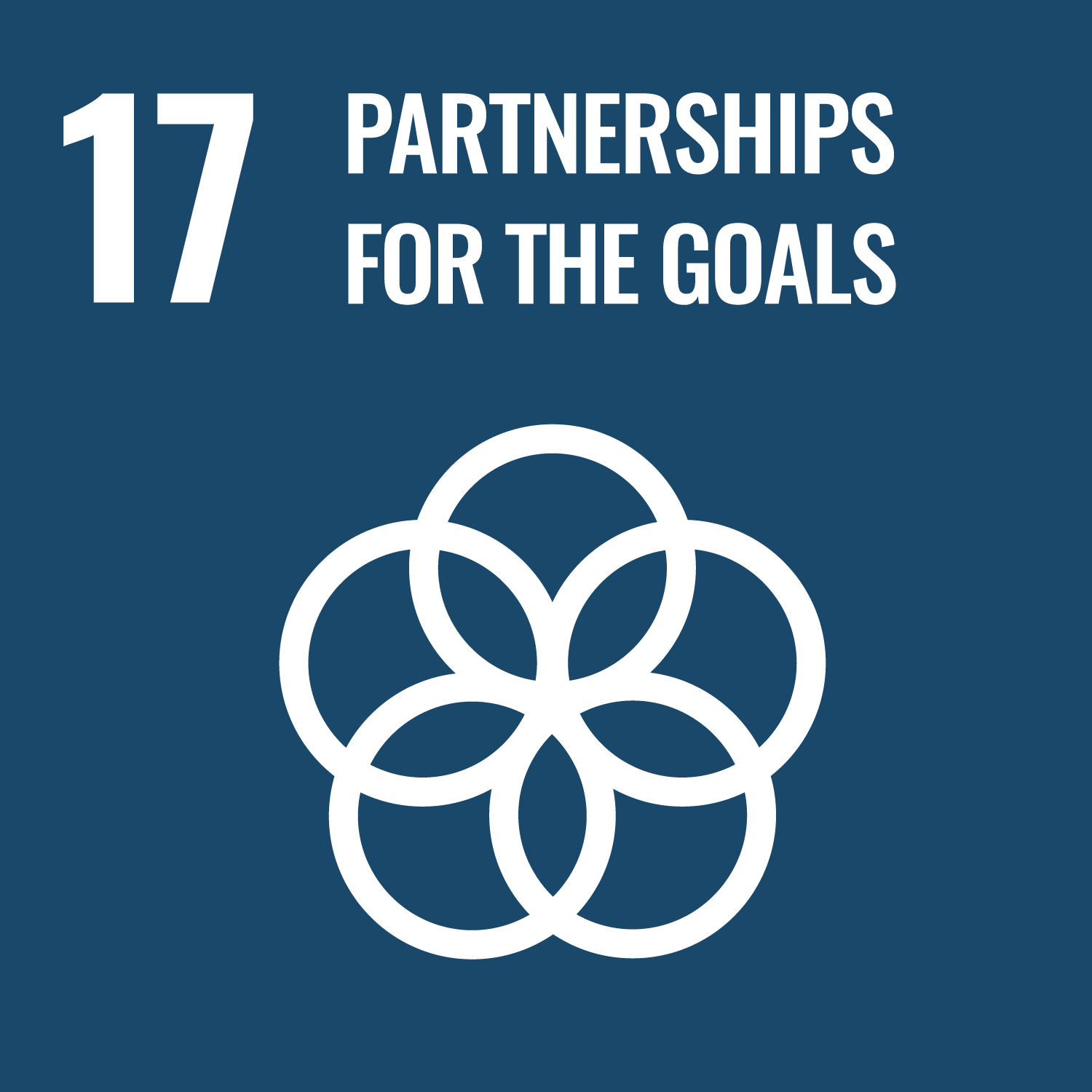}
    \caption{Examples of Sustainable Development Goals (SDGs) that converge with Astrotourism. Image Credit: \href{https://sdgs.un.org/goals}{United Nations Sustainable Development Goals}}\label{sdgs}
\end{figure}

Given its potential, the OAD has sought to ensure that astrotourism is not merely a luxury niche but a development tool accessible to communities around the world.

\section{Openness, Accessibility, and Evolving Resources}
At the heart of this initiative is a commitment to openness and accessibility. The resources discussed in this paper are freely available and easy to use by a wide range of individuals and communities, regardless of background, location, or expertise. We believe that knowledge and tools related to astrotourism should be shared as widely as possible to foster inclusive and locally grounded innovation.

These resources are not static. They are living resources, designed to grow and improve over time through collaboration and ongoing feedback from users. We encourage you to adapt, expand, and share them within your own context, and to contribute your insights and experiences to strengthen the collective impact.

To support this philosophy, all our resources are released under a \href{https://creativecommons.org/licenses/by/4.0/legalcode}{Creative Commons licence}, which clearly outlines permissions for reuse, adaptation, and distribution. This approach ensures that the materials remain open and beneficial to all, while also recognising the contributions of those who build on them.

\textbf{Potential Use Cases}\\
These resources are designed to be adaptable across different geographies and cultural contexts, making them relevant for a wide range of users. They can be applied in various ways, such as launching a night-sky tourism business in rural or peri-urban areas, developing local or national dark-sky tourism strategies, and blending storytelling, science, and sustainability in community-based tourism initiatives.

\section{Overview of OAD Astrotourism Resources}
By broadening the scope of astrotourism activities, the OAD aims to make it easier for communities to get involved. No longer is it necessary to invest in expensive telescopes or astronomy expertise. Instead, the focus is on creating welcoming and immersive experiences that celebrate the night sky as a shared resource.

This community-centric approach not only benefits local economies but also preserves invaluable cultural heritage and fosters a deeper connection between people and the natural world. The OAD has put together resources that aim to facilitate the bridge between astronomy and community-led experiences in meaningful ways. 

Our resources are built to leverage the strengths of various stakeholder groups rather than focus solely on the gaps. The resources were created with the following target audiences in mind:
\begin{itemize}
\item \textbf{Individuals (Resource 1)}\\
This resource is targeted at inhabitants of rural, socioeconomically underdeveloped areas. Their strengths are that they generally live in areas with abundant, free dark skies, and possess distinct cultural elements like food, rituals, and way of life.
\item \textbf{Existing Businesses (Resource 2)}\\
This resource is targeted at existing tourism businesses who want to implement astrotourism. Their strengths are existing tourism skills, infrastructure, and established logistical elements of experience.
\item \textbf{Communities around Observatories (Resource 3)}\\
This resource is targeted at communities around observatories. Observatories can help surrounding community inhabitants/organisations create experiences to complement the observatory's offerings. The observatory's strengths are their infrastructure, connection or proximity to community, technical expertise and knowledge, links to government. 
\end{itemize}

\begin{figure}[H]
    \centering
    \includegraphics[width=0.3\textwidth]{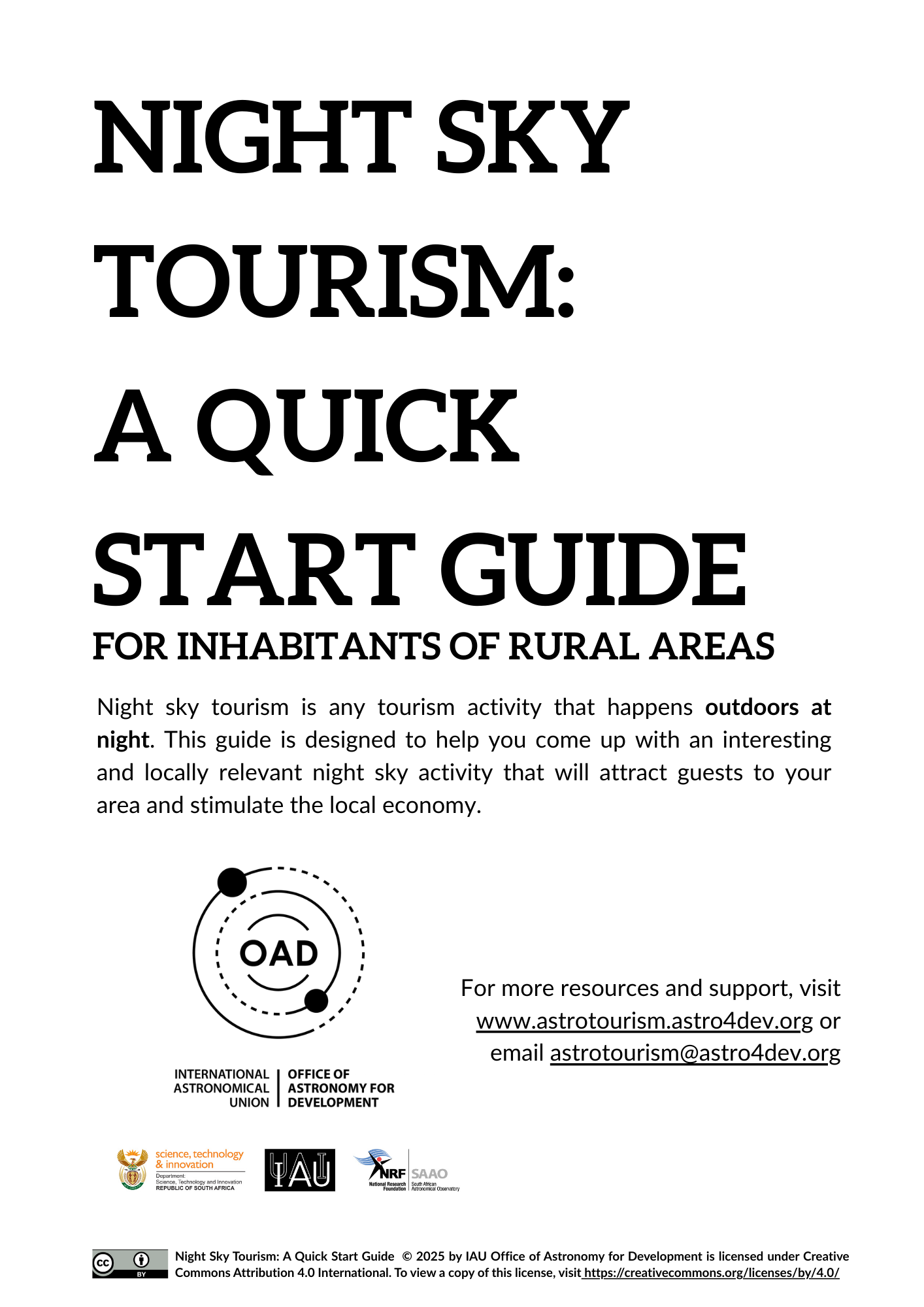}
    \includegraphics[width=0.3\textwidth]{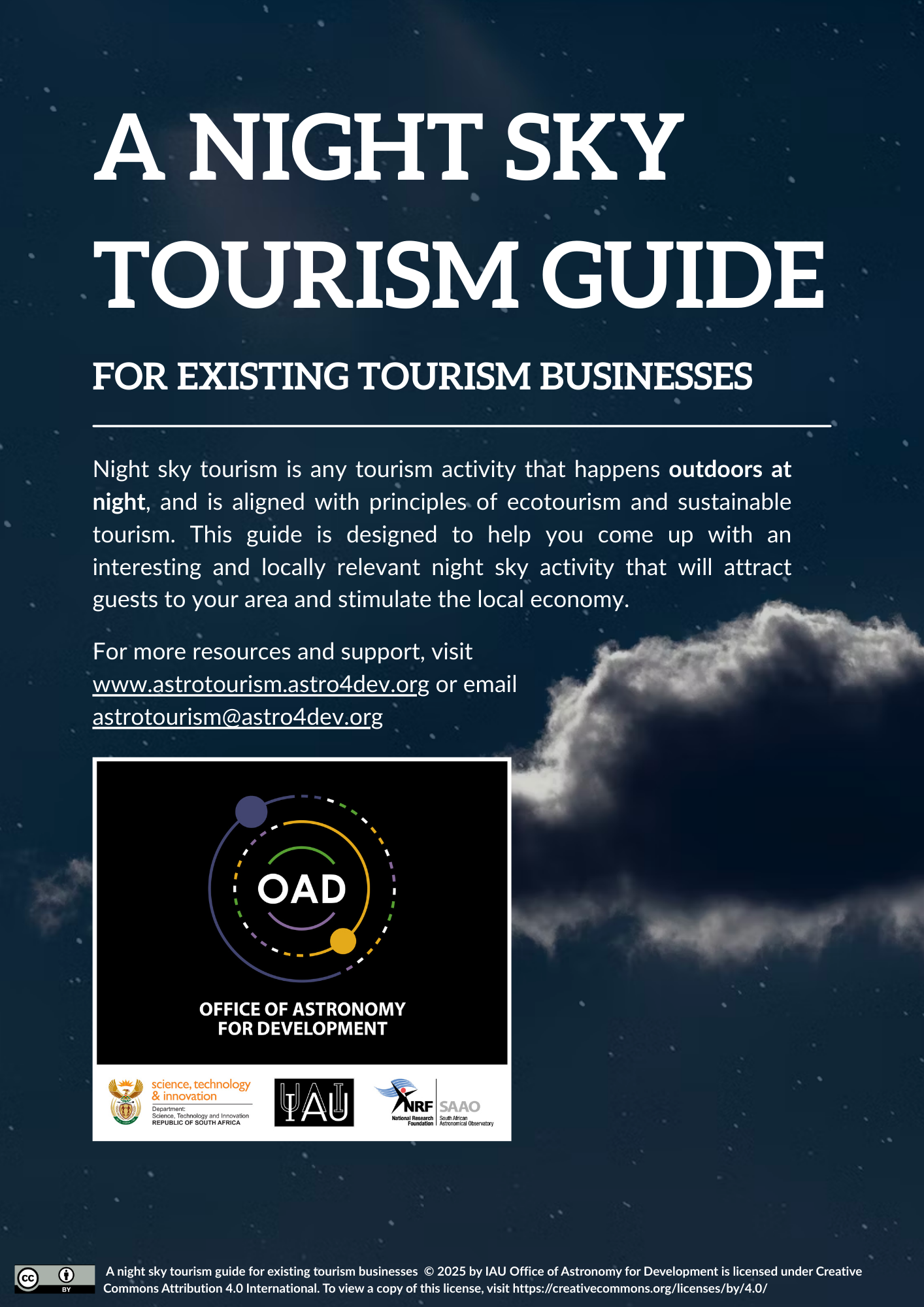}
    \includegraphics[width=0.3\textwidth]{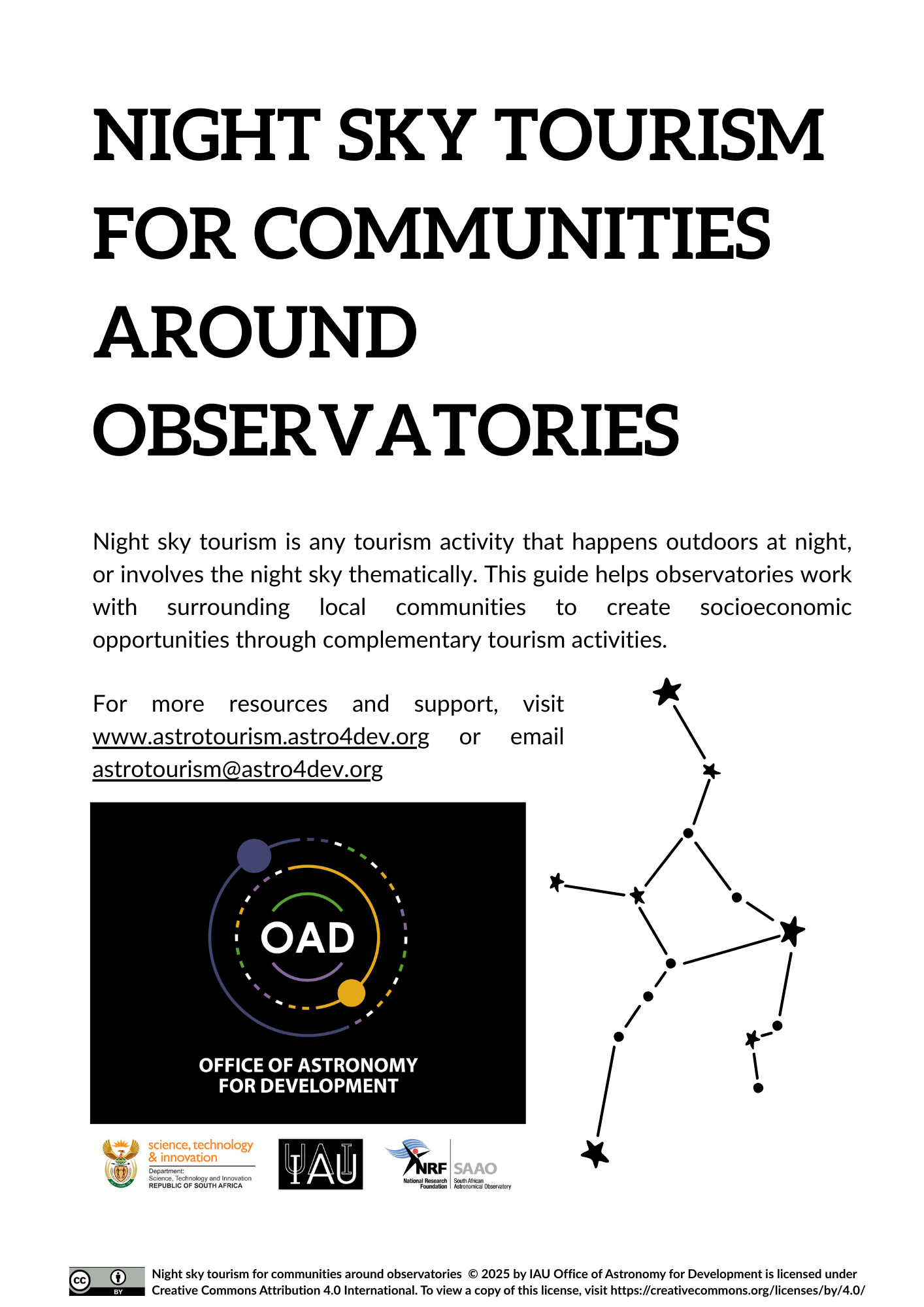}
    \caption{Astrotourism resources created by the OAD to support individuals, existing tourism businesses, and communities around observatories in their entry into astrotourism. Available on the \href{https://astrotourism.astro4dev.org/resources}{OAD Astrotourism Portal} \cite{astrotourism}}\label{resources}
\end{figure}

\subsection{Navigating the Resources}
The resources are designed to complement each other. While each one targets a specific audience, parts of them can be integrated for broader use. Each resource also cross-references the others in relevant sections to ensure coherence. To guide users, a flowchart aimed at helping individuals and organisations identify the most suitable night-sky tourism resources to get started is attached at the end of the paper.

Below a more detailed description of each resource is provided:

\subsection{Resource 1 - Night Sky Tourism: A Quick Start Guide}\label{resource 1}
The "Night Sky Tourism: A Quick Start Guide" aims to help rural communities create engaging night sky activities to attract tourists and boost the local economy. This resource targets inhabitants of rural, socioeconomically underdeveloped areas. The resource aims to highlight their strengths and how they can use these strengths to their advantage. The Quick Start Guide is a short crash course on creating a night sky tourism product for rural areas. It is designed for visual emphasis and affordable printing.

The guide outlines eight key steps that are briefly summarised below:
\begin{itemize}
\item \textbf{Choice of Nighttime Activity:}
Select an outdoor activity that includes the night sky as a backdrop. Choose from traditional dinners under the stars, cultural performances, art activities, wellness practices, or nighttime walks. Consider existing skills and resources to minimise the need for additional materials.
\item \textbf{Use of Local Crafts and Objects:}
Incorporate local crafts such as pottery, basketry, textiles, and beadwork to add authenticity. Offer these as paid activities for visitors to participate in, or sell pre-made products as an additional source of income.
\item \textbf{Cultural Astronomy:}
Integrate cultural astronomy into astrotourism offerings to create cultural immersion and shared experiences that bridge social and economic divides. Research and include unique cultural perspectives on the Sun, Moon, stars, constellations, and special celestial events like meteor showers and eclipses to enhance the experience.
\item \textbf{Logistics of the Activity:}
Ensure the location is safe and that all necessary permissions are in place. Consider accessibility for all guests and prepare for emergencies. Set prices that are both profitable and affordable.
\item \textbf{Test Run:}
Conduct a trial run with friends and family to gather feedback, identify potential issues, and make improvements.
\item \textbf{Extra Tips: Environmental Sustainability \& Protecting the Night Sky:}
There is an emphasis on responsible tourism practices. Avoid littering, limit participant numbers, minimise noise, and use eco-friendly lighting such as red lights, candles, firelight, or shielded fixtures.
\item \textbf{Spreading the Word: Marketing \& Promotion:}
Promote the experience to attract customers. Use social media, local events, media outreach, websites, email lists, referral programmes, community boards, and partnerships with local businesses and tourism agencies. Consider platforms like Airbnb Experiences, Viator, and Eventbrite.
\item \textbf{Next Steps:}
Develop a business plan using the Business Model Canvas from the "Night Sky Tourism for Communities around Observatories" resource (see Section \ref{resource 3}).
\end{itemize}
To accompany this guide, a workbook was created to guide users through the Night Sky Tourism Quick Guide, going through the workbook will leave the user with a draft of a workable plan that is ready for implementation \cite{resource1}.

\subsection{Resource 2 - A Night Sky Tourism Guide}
The "A Night Sky Tourism Guide For Existing Businesses" aims to help existing tourism businesses develop night sky tourism activities that are engaging, culturally relevant, and sustainable. The guide aims to bridge the transition into astrotourism by providing practical insights and step-by-step instructions for integrating night sky experiences into existing tourism offerings. The goal is to empower businesses to introduce unique, memorable, and sustainable night sky tourism products that appeal to travelers seeking authentic and meaningful experiences.

Key takeaways from the guide include:
\begin{itemize}
\item \textbf{Understanding Night Sky Tourism:} The guide provides a definition of night sky tourism as any outdoor night activity aligned with ecotourism and sustainable tourism principles. It includes astrotourism (science-focused activities such as stargazing with telescopes) and celestial tourism (observing celestial events like meteor showers). It also broadens to include activities that use the night sky as a backdrop, such as food experiences, wellness practices, and cultural exploration.
\item \textbf{Market Opportunity:} Statistics indicate that 8 out of 10 people live under light-polluted skies \cite{fabio}, making areas with access to dark skies in possession of unique opportunities for tourism offerings. Additionally, nighttime activities also offer less competition than daytime activities.
\item \textbf{SOAR Analysis:} The guide recommends a Strengths, Opportunities, Aspirations, and Results (SOAR) analysis to leverage existing business assets (infrastructure, local expertise, customer base) and capitalise on opportunities like low competition and growing interest in stargazing.
\item \textbf{Meaningful Experiences:} The guide places emphasis on creating meaningful tourism experiences by incorporating altruism (contributing to local communities), pleasure (beauty, relaxation), and connection (authenticity, immersion, sense of belonging).
\item \textbf{Experience Design:}  Users of the guide are provided with activities for designing experiences. Some of the key themes that are considered in experience design are listed below:
\begin{itemize}
\item \textbf{Cohesion:} The first step involves analysing existing themes in current offerings such as wildlife, heritage and wellness. Once the existing themes are identified, designing complementary night sky activities will build a cohesive narrative for visitors.
\item \textbf{Resource Analysis:} There are three key resources that can be integrated into experience, the following resources can be considered to enhance the experience: 1) environmental conditions such as the weather and especially the light pollution levels, 2) ecological assets such as natural landscapes, biodiversity and geological features and 3) cultural resources such as local traditions, heritage sites, art, and cultural astronomy.
\item \textbf{Designing the Experience:} Special focus should be on theme integration, storytelling, tying the experience to local landscape and culture, and creating immersive experiences through physical cultural touchpoints, engaging senses with natural darkness, incorporating local food, and including hands-on activities.
\item \textbf{Safety and Comfort:} Crucial considerations for night activities include temperature, ecological threats, visibility/lighting (recommending red light), and emergency planning.
\item \textbf{Astronomy Resources:} For astronomy-focused experiences, it is recommended to use tools like Stellarium (for planning and learning), \href{https://www.timeanddate.com/astronomy/}{timeanddate.com/astronomy} (for celestial events), and equipment like binoculars and telescopes, as well as mobile stargazing apps and printable star maps.
\item \textbf{Dark Sky Friendly Lighting:} The guide promotes responsible outdoor lighting practices to protect the night sky and nocturnal environment, adhering to \href{https://darksky.org/news/darksky-international-releases-guiding-principles-for-responsible-astrotourism/}{DarkSky International's Five Lighting Principles} (useful, targeted, low level, controlled, warm-coloured) that were endorsed by the OAD \cite{darksky}.
\end{itemize}
\end{itemize}

This resource guides users in expanding their offerings from daytime to include nighttime experiences. It features activities that involve users in co-designing the experience \cite{resource2}.

\subsection{Resource 3 - Night Sky Tourism for Communities Around Observatories}\label{resource 3}
The "Night Sky Tourism for Communities Around Observatories" resource targets communities around observatories. Observatories are in a good position to co-create experiences with rural inhabitants in surrounding communities. Their strengths are the ability to provide extended support to rural inhabitants to help them create a night sky experience. The resource aims to guide the user into curating a variety of complementary activities that can take place during an observatory's downtime, integrating astronomy to generate additional revenue for local communities. It is used along with the "Night Sky Tourism: A Quick Start Guide" resource but existing tourism businesses located around observatories can also use the guide.

There is special emphasis on communities and observatories working together to co-create and identify complementary activities that showcase local cultural and ecological assets. These experiences should be co-created between community inhabitants and observatories to enhance the observatory experience and provide recreational activities, food, and accommodation during observatory downtimes.

The guide is divided into three main sections:
\begin{itemize}
\item \textbf{Section 1: Itinerary \& Identifying Themes:} This section is designed for observatory staff, it focuses on identifying observatory downtimes and gaps in services that can be filled with cultural, ecological, and family-friendly activities. It also guides observatories in identifying and leveraging their central themes such as celestial objects and/or astronomers, that contribute towards creating cohesive visitor experiences.
\item \textbf{Section 2: Designing the Experience:} This section is designed to support organisations and community members. It outlines how to allocate business activities to available time slots based on the observatory’s downtime. Individuals or groups are categorised into four areas—accommodation, food vendors, experiences, and items for sale—and can then be assigned to the relevant time slots. Users are then directed to a separate resource, "Night Sky Tourism: A Quick Start Guide" (Resource 1 - see section \ref{resource 1}), to further develop these business ideas.
\item \textbf{Section 3: The Business Model Canvas:} This section introduces the Business Model Canvas as a strategic tool to develop and document business models. It explains the nine essential building blocks: Key Partners, Key Activities, Key Resources, Value Propositions, Customer Relationships, Channels, Customer Segments, Cost Structure, and Revenue Streams. An additional "Astronomy USP (Unique Selling Proposition)" section is included to show how businesses can tie in night sky tourism themes. The guide provides sample and empty canvases for the various categories along with examples for filling them out.
\end{itemize}
The guide concludes by highlighting the importance of collaboration between observatories and community members to enhance economic opportunities and enrich the overall visitor experience \cite{resource3}. 

\section{Summary and Conclusion}
These resources aim to encourage the creation of authentic nature-based night sky experiences in rural areas that celebrate their strengths such as natural heritage and dark skies due to their remote location, and rich cultural heritage that can be leveraged to create unique and meaningful encounters for tourists. By leveraging their strengths to create simple and relevant tourism products themed around the night sky, the aim is to economically empower them, incentivise the protection of dark skies, and break down social barriers between visitors and practitioners.

Astrotourism, when supported by inclusive and practical tools, has the potential to become a powerful instrument for sustainable development. The IAU Office of Astronomy for Development (OAD) provides resources that enable communities and individuals around the world to harness the power of astronomy to inspire, educate, and drive positive change.

\textbf{Reflections and Future Plans}\\
The OAD’s astrotourism resources are living documents that are freely available, accessible, and specifically designed for low-resource environments. They are flexible and can be adapted to different cultural contexts, making them suitable for diverse communities around the world. Looking ahead, recommended enhancements include expanding the resources into more languages, requiring volunteers who are willing to assist with translations. Additionally, there is a need for more regionally diverse case studies, individuals and organisations are invited to share their experiences and best practices in astrotourism. Another key area for development is the creation of frameworks to monitor and evaluate the impact of these initiatives.

If you have any suggestions on how we can enhance these resources to better serve the global astrotourism community, we value your input and would love to hear from you. Whether you have ideas to improve their content, make them more accessible, share feedback based on your experience using them, or if you are interested in helping translate them into other languages.

Visit our website \href{https://www.astro4dev.org/astrotourism}{https://www.astro4dev.org/astrotourism} or feel free to reach out to us at \href{astrotourism@astro4dev.org}{astrotourism@astro4dev.org}

\section*{Acknowledgements}
The IAU Office of Astronomy for Development acknowledges Samyukta Manikumar, for conceptualising and creating the resources, along with the collaborators and contributors to the Astrotourism Resources. We also thank the communities and practitioners who have tested and refined these resources through real-world application.

\includepdf[pages=-]{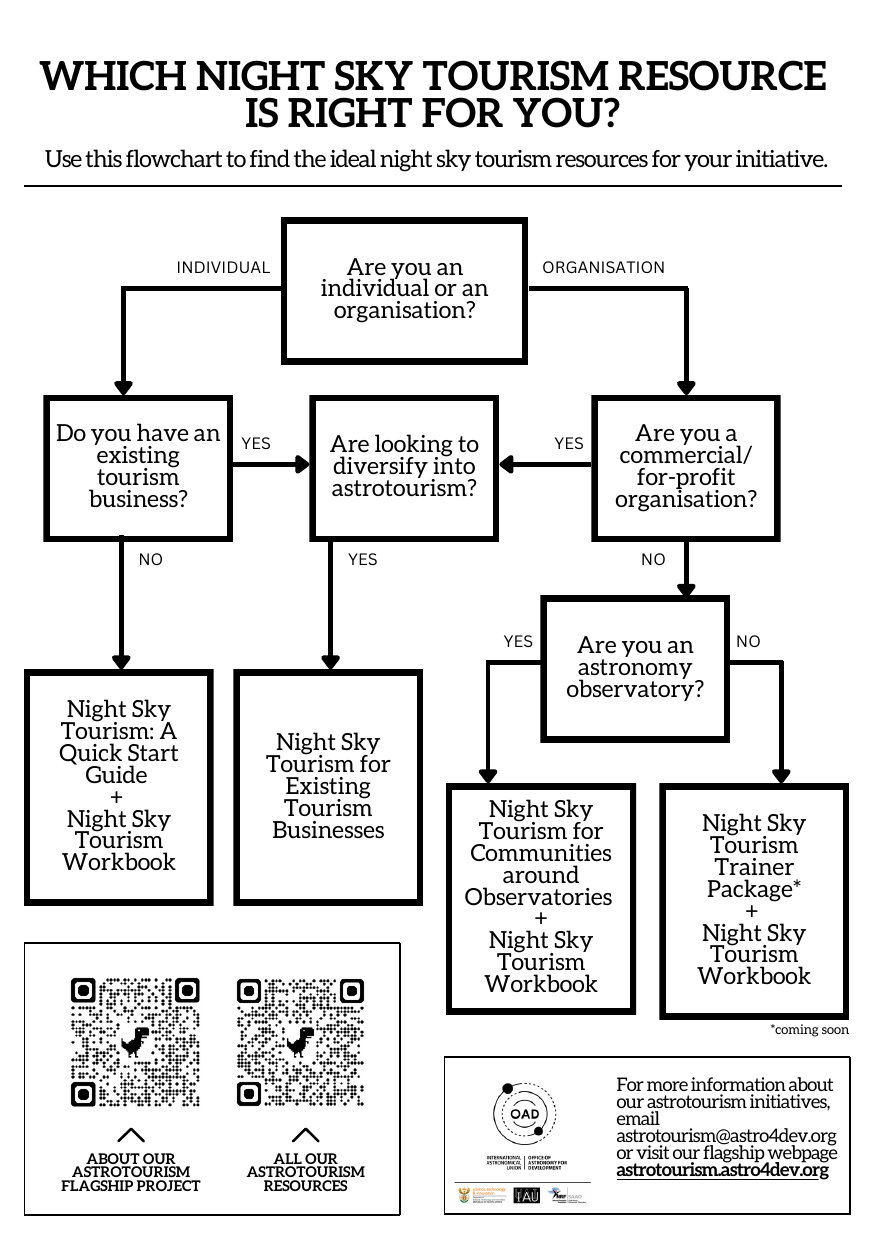}

\end{document}